\documentclass{ar-1col-S2O}
\usepackage[numbers]{natbib}
\usepackage{url}
\newcommand{\gtorder}{\mathrel{\raise.3ex\hbox{$>$}\mkern-14mu
            \lower0.6ex\hbox{$\sim$}}}
\newcommand{\ltorder}{\mathrel{\raise.3ex\hbox{$<$}\mkern-14mu
            \lower0.6ex\hbox{$\sim$}}}

\jname{Annu. Rev. Nucl. Part. Sci.}
\jvol{76}
\jyear{2026}
\doi{10.1146/annurev-nucl-100824-043815}

\begin{document}

% Page header
\markboth{M. Coleman Miller}{Neutron Star Radii from NICER}

% Title
\title{What We Have Learned from NICER About Neutron Star Radii}

%Authors, affiliations address.
\author{M. Coleman Miller,$^1$
\affil{$^1$University of Maryland, Department of Astronomy and Joint Space-Science Institute, College Park, MD 20742--2421, USA}}

%Abstract
\begin{abstract}

The cores of neutron stars have a combination of density, temperature, and neutron-proton asymmetry that cannot be replicated in laboratories or understood using first-principles quantum chromodynamics.  Thus, observations of neutron stars are necessary to understand this regime.  In particular, precise and accurate measurements of neutron star radii are highly informative about cold, catalyzed matter at a few times nuclear saturation density.  X-ray observations with NASA's Neutron star Interior Composition Explorer (NICER) have yielded radius measurements for a few nonaccreting neutron stars, which have advanced substantially our understanding of dense matter.  In this review we discuss the NICER radius measurements and their implications and demonstrate that the systematic errors thus far explored in NICER analyses have at most a minor effect on the inferred radii of neutron stars.

\end{abstract}

%Keywords, etc.
\maketitle

%Table of Contents
\tableofcontents

% Heading 1
\section{INTRODUCTION}

Neutron stars are one outcome of the collapse of the core of a massive star (with an initial mass more than $\sim 8~M_\odot$, where $M_\odot\approx 2\times 10^{33}$~g is the mass of the Sun).  Observationally and from stellar evolution theory, neutron stars have gravitational masses (the mass we would measure using a distant orbiting satellite and Kepler's laws; hereafter, we call this simply ``mass") from $\sim 1.2~M_\odot$ to somewhat above $2~M_\odot$ (e.g., \cite{2012ARNPS..62..485L}), and circumferential equatorial radii (the radius we would obtain by measuring the equatorial circumference of the star and dividing by $2\pi$; hereafter, we call this simply ``radius") that, prior to any observations, were expected to be in the range $\sim 8-16$~km \cite{1983bhwd.book.....S}.  Hence the average density of a neutron star is likely to exceed nuclear saturation density ($\rho_s\approx 2.6\times 10^{14}$~g~cm$^{-3}$, the maximum density achieved in atomic nuclei without external confining pressure \cite{2012ARNPS..62..485L}), and the central densities of neutron stars can reach several times $\rho_s$.  

In addition, the core temperatures ($T_c\ltorder 10^{10}$~K for stars $>1$~year old \cite{2001MNRAS.324..725G}) of neutron stars are orders of magnitude smaller than their Fermi temperatures $T_F\sim 10^{12}$~K.  Thus their temperatures can usually be neglected in calculations of their structure, and in beta equilibrium there are $\sim 10$ or more neutrons per proton, so the cores are highly neutron-proton asymmetric.  This combination of high density, relatively low temperature, and large neutron-proton asymmetry cannot be replicated in laboratories.  Moreover, because the baryons in neutron stars have a nonzero net density, first-principles calculations of the equilibrium properties of such matter from quantum chromodynamics (QCD) are hampered by the so-called fermion sign problem \cite{2005PhRvL..94q0201T,2010IJMPA..25...53H}.  As a result, observations of neutron stars are necessary to obtain insight into this state of matter, which in some sense provides a bridge between conditions accessible to experiment and the theoretically simple realm of extremely high densities, where QCD becomes perturbative.

Here we focus on the measurements of neutron star radii and, in particular, the contributions made by analyses of data from NASA's Neutron star Interior Composition Explorer \cite{2016SPIE.9905E..1HG}.  Radii, if measured with sufficient precision and if not plagued by substantial systematic errors, provide particularly useful information for the construction of models of neutron star core matter.  In Section~\ref{sec:NSprop} we give a brief review of how we compute neutron star properties such as the maximum mass, the radius as a function of mass, the tidal deformability, and the moment of inertia given an equation of state of dense matter.  In Section~\ref{sec:NSnonradius} we discuss neutron star measurements other than of the radius.  In Section~\ref{sec:NSradpreNICER} we review some of the work prior to the launch of NICER that attempted to constrain neutron star radii, and the large systematic errors that were discovered in these approaches.  In Section~\ref{sec:NICER} we give a brief overview of the NICER mission itself and the surface emission models we use, followed by a description in Section~\ref{sec:systematic} of potential systematic errors and why these are not problematic, and a summary of the results from NICER analyses in Section~\ref{sec:results}.  We link the results from NICER and other astronomical observations to nuclear physics in Section~\ref{sec:nuclear}, and present our conclusions in Section~\ref{sec:summary}.

\section{CALCULATION OF NEUTRON STAR PROPERTIES FROM THE EQUATION OF STATE}
\label{sec:NSprop}

At the densities and temperatures of neutron star cores the dominant degrees of freedom are unknown.  Is the core predominantly neutrons, protons, electrons, and muons?  Do hyperons play an important role?  Is there a transition to free quarks and gluons?  Structural measurements of neutron stars (such as measurements of their radii) do not touch directly on the degrees of freedom.  Instead, for the neutron stars that we can observe, what matters is the equation of state (EOS).  The EOS, in general, can be represented as the pressure as a function of other quantities, which could include the density, temperature, composition, and so on.  For neutron star cores, as we indicated earlier the temperature can be neglected.  It is also assumed (probably correctly, although this has not been proven rigorously) that neutron star core matter at a given density is equilibrated at that density and thus we do not have to contemplate different compositions.  Given those assumptions, the same EOS describes all neutron stars; some neutron stars have larger central densities than others and thus could access different degrees of freedom, but at a given density, we assume that any neutron star will have the same pressure.  Then the pressure $p$ is a function of only the energy density $\epsilon$: $p=p(\epsilon)$.  

Given an assumed EOS, for a nonrotating and thus spherical star the equation of hydrostatic equilibrium in general relativity is the Tolman-Oppenheimer-Volkoff (TOV) equation \cite{1939PhRv...55..364T,1939PhRv...55..374O}:
\begin{equation}
\frac{dp}{dR}=-\frac{(\epsilon+p)(M(<R)+4\pi R^3p/c^2)}{R(c^2R/G-2M(<R))}\; .
\end{equation}
Here $R$ is the radius, $M(<R)$ is the mass inside $R$, $c$ is the speed of light in a vacuum, and $G$ is Newton's gravitational constant.  Starting from a given EOS and central energy density $\epsilon_c$, we can integrate outward to some threshold pressure (which is small but finite to avoid numerical instabilities) and can thus find the mass and radius associated with that central energy density and EOS.  A property of the TOV equation not shared by the Newtonian equation of hydrostatic equilibrium is that for a given EOS there is a maximum stable mass.  

One can also calculate, with much more complicated formulae, the moment of inertia \cite{1967ApJ...150.1005H}, the rotational quadrupole moment \cite{1967ApJ...150.1005H,1968ApJ...153..807H}, and the gravitational tidal deformability \cite{2008ApJ...677.1216H,2009ApJ...697..964H} of a slowly rotating neutron star constructed from a given EOS with a specified central energy density.

The restriction to nonrotating or slowly rotating neutron stars needs to be compared quantitatively to the rotation frequencies of the stars that we observe.  As an approximate comparison, suppose that a typical neutron star has a mass $M=1.4~M_\odot$ and a radius $R=12$~km.  The Keplerian rotational frequency at the stellar surface is $\nu_K(R)=(GM/R^3)^{1/2}/(2\pi)\approx 1600$~Hz.  Symmetry arguments indicate that the deviation of the shape of a star from a sphere scales as the square of the rotational frequency.  That is, roughly, we could guess that for a star spinning at rotational frequency $\nu$, $(R_{\rm eq}-R_{\rm pole})/R_{\rm eq}\sim (\nu/\nu_K)^2$, where $R_{\rm eq}$ is the equatorial radius and $R_{\rm pole}$ is the polar radius; this fractional difference is actually multiplied by $\approx 1/2$ for real stars \cite{2007ApJ...663.1244M}.  The neutron stars observed with NICER have rotation frequencies $\nu\sim 200-300$~Hz, which means that we expect deviations from sphericity on the order of a few percent.  The result is that rotation has a small effect on the neutron stars studied with NICER, except that we do take the resulting oblateness into account when we calculate the rotation-phase-dependent spectra that we expect from spots on the surface \cite{2007ApJ...663.1244M,2014ApJ...791...78A,2019ApJ...887L..26B,2021PhRvD.103f3038S}.  Similarly, for these frequencies frame-dragging in the external spacetime changes the waveforms by amounts much smaller than the precision of NICER data (see, e.g., Figure~2 in \cite{2019ApJ...887L..26B}).

\section{NEUTRON STAR MEASUREMENTS OTHER THAN THAT OF THE RADIUS}
\label{sec:NSnonradius}

To provide context for the NICER measurements, in this section we discuss neutron star measurements other than the radius, and in Section~\ref{sec:NSradpreNICER} we summarize pre-NICER attempts to measure the radius and our current understanding of the potential systematic errors in such methods.

\subsection{Neutron star mass measurements}
\label{sec:NSmass}

As mentioned in Section~\ref{sec:NSprop}, a given EOS implies a maximum mass for a nonrotating neutron star.  Rotating neutron stars add centrifugal support, but as indicated above, the impact is minimal for most stars (although for rotation frequencies $\gtorder 600$~Hz the corrections become more substantial; see, e.g., Figure~13 from \cite{2019ApJ...887L..24M} and recall that deviations in the radius of a star scale as the square of the rotational frequency).  Since an EOS has a maximum mass, this means that, modulo measurement uncertainties, if we establish that even one neutron star has a mass greater than the maximum allowed for a given EOS, that EOS cannot be correct.  Thus measurements of the masses of high-mass neutron stars provide strong constraints on the EOS.

The most precise such measurements are made for binaries containing neutron stars that are observed as radio-emitting pulsars.  Pulsars are excellent natural clocks, so the imprint of Doppler shifts on the pulsar signal due to binary motion can be detected easily.  In addition to the Newtonian effects, there are general relativistic contributions to the timing (collectively called post-Keplerian effects) which allow for an overdetermination of the system: in addition to the masses of both binary components, their separation or orbital period, their orbital eccentricity, and the inclination of the orbit to our line of sight, the underlying theory (general relativity) can be tested precisely for neutron star binary systems \cite{2009arXiv0907.3219F}.  

One of the most important of the post-Keplerian effects is Shapiro delay \cite{1964PhRvL..13..789S}.  The delay in arrival of photons from a pulsar is caused by the motion of those photons through the gravitational field of the companion.  The orbit modulates that delay, in a way that depends on the orientation of the orbit relative to our line of sight: for a face-on orbit there is no modulation, whereas for an edge-on orbit the delay is maximally modulated.  The magnitude of the delay depends on the orientation and on the mass of the companion.  Thus measurement of the Shapiro delay yields two quantities, the ``range" (related to the mass of the companion) and the ``shape" (related to the orientation of the binary relative to our line of sight).  Combined with the information from Doppler modulation (which gives the binary eccentricity, the orbital period, and the orbital velocity projected onto our line of sight), the Shapiro delay therefore yields enough information to solve for the eccentricity, orbital separation, orbital inclination, and both masses.  Critically for the robustness of the effect, the Shapiro delay does not depend on the structure of the companion (e.g., whether it is an ordinary star or a neutron star) as long as the propagating photons do not intersect the companion or its magnetosphere.  For orbits with intermediate orientations relative to our line of sight there is some degeneracy between orbital eccentricity and the Shapiro delay, but for nearly edge-on orbits the delay has a cusp near the pulsar-companion-Earth passage that is clearly distinguishable from eccentricity.

Of the pulsars which have had their masses measured using Shapiro delay, two stand out as having especially high masses: PSR~J1614$-$2230 ($M=1.937\pm 0.014~M_\odot$ \cite{2010Natur.467.1081D,2023ApJ...951L...9A}) and PSR~J0740$+$6620 ($M=2.08\pm 0.07~M_\odot$ \cite{2020NatAs...4...72C,2021ApJ...915L..12F}; this is one of the pulsars observed with NICER).  These two $\sim 2~M_\odot$ neutron stars place strong constraints on the high-density EOS, and rule out strong low-density phase transitions in the core.

High pulsar masses have been suggested by other data as well.  An example of the challenges involved in other techniques is the case of PSR~J0348$+$0432.  This pulsar orbits a white dwarf, and although the orbital orientation is not favorable for Shapiro delay, gravitationally redshifted absorption lines were seen from the white dwarf.  The observed energies of these lines are modulated by the orbit, which combined with the modulation of the observed pulsar frequency gives the mass ratio between the neutron star and the white dwarf.  Modeling of the gravitationally redshifted lines yields the white dwarf mass, which combined with the mass ratio gives the neutron star mass.  Based on this modeling, \cite{2013Sci...340..448A} reported a mass of $M=2.01\pm 0.04~M_\odot$ for PSR~J0348$+$0432.  However, continued observation of the system found evidence of decay of the orbit due to gravitational radiation, and the rate of decay implies a substantially lower mass of $M=1.806\pm 0.037~M_\odot$ \cite{2025ApJ...983L..20S}.  

Another recent report of a high neutron star mass concerns the ``black widow" pulsar PSR~J0952$-$0607.  Black widows are pulsars whose radiation is at least partially destroying their very low-mass companions.  These systems are tricky to model, but the best estimates of the pulsar masses have typically been quite high, e.g., $2.68\pm 0.14~M_\odot$ for PSR~J1311$-$3430 \cite{2015ApJ...804..115R}.  However, the fits have normally had substantial residuals (see Figure~9 in \cite{2015ApJ...804..115R} for an example), which have led the authors of these papers to urge caution in their interpretation.  In contrast to its fellow black widows, the best fit to the PSR~J0952$-$0607 light curve does not have obvious patterns in its residuals (see Figure~1 in \cite{2022ApJ...934L..17R}).  Thus its reported mass of $M=2.35\pm 0.11~M_\odot$ \cite{2026ApJ...996..101R} is more reliable than the masses inferred from other black widows.  However, there are still substantial assumptions that go into the modeling, such as the nature of the interaction of the pulsar wind with the companion, so a conservative approach would be to use only the masses of PSR~J1614$-$2230 and PSR~J0740$+$6620 to place lower limits on the maximum mass of a neutron star, and thus lower limits on the pressure at the relevant densities of a few times nuclear saturation density.

\subsection{GW170817 and its afterglow}
\label{sec:GW170817}

On 17 August 2017, ground-based gravitational-wave detectors observed a signal consistent with the coalescence of two neutron stars \cite{2017PhRvL.119p1101A}.  Roughly 1.7 seconds after the merger, a gamma-ray burst was seen, and later observations revealed emission across the whole electromagnetic spectrum \cite{2017ApJ...848L..12A}.  This event was transformative in numerous ways: for example, it confirmed the suspected association between neutron star mergers and short gamma-ray bursts (at least in one case!), it showed the predicted ``kilonova" signal produced by radioactive decay of the heavy elements synthesized in the outflow, and it allowed new constraints to be placed on the EOS \cite{2018PhRvL.121p1101A,2018PhRvL.121i1102D}.

The EOS constraints emerge in the following way.  When the separation between two orbiting neutron stars is many times their radii, the orbit and the resulting gravitational radiation are essentially the same as they would be if the stars were points.  However, when the stars are closer to each other, their mutual tidal gravitational field distorts the stars out of their nearly-spherical equilibrium.  This takes energy, which is drawn from the orbit, and that combined with the different rate of gravitational radiation due to the non-spherical shape increases the rate at which the stars spiral toward each other.  These deviations from point-mass behavior are much stronger when the stars are closer to each other.  This leaves an imprint on the gravitational waveform.  The more easily a star is distorted by tidal fields, the stronger is the effect on the waveform.  This, therefore, provides a new probe of the structure of neutron stars.

The signature of tidal deformation was at best weakly detected \cite{2017PhRvL.119p1101A}, so the observation yields an upper limit to the tidal deformability of the stars.  Although tidal deformability depends on the full structure of the star and not just its radius, within broad categories of EOS one can roughly say that at 90\% credibility the stellar radius is $<13.6$~km for these stars \cite{2018PhRvL.120q2703A,2018ApJ...857L..23R}, which have estimated masses $\approx 1.35~M_\odot$ \cite{2017PhRvL.119p1101A}.  This, in turn, places an upper limit on the pressure at the relevant densities (perhaps twice nuclear saturation density), so it complements the constraints from the existence of high-mass neutron stars.

One point to emphasize is that higher-mass neutron stars are much less easy to deform tidally than lower-mass neutron stars.  Normalized appropriately, the tidal deformability is expected to scale as roughly $M^{-6}$ with the mass $M$ of the star (assuming that the radius is roughly constant with mass; see \cite{2018PhRvL.121i1102D}).  Given that even the strong and apparently rare event GW170817 (no similar event has been seen since) only provided an upper limit of the deformability of a $\approx 1.35~M_\odot$ star, more massive neutron stars such as PSR~J0740$+$6620, which have substantially higher central densities than lower-mass stars, must be constrained in other ways.  As discussed below, NICER observations provide that window.

For completeness, we also note that the afterglow of GW170817 has been analyzed to provide more model-dependent constraints on the EOS.  The merger between two neutron stars is not clean; some hundredths to tenths of a solar mass are ejected in the process, even if a black hole is formed.  Several groups conclude from their numerical modeling that the electromagnetic and gravitational-wave data on GR170817 suggest an upper limit on the maximum mass of a nonrotating star, in the vicinity of $2.1-2.3~M_\odot$ \cite{2017ApJ...850L..19M,2018PhRvL.120z1103M,2018ApJ...852L..25R,2018PhRvD..97b1501R}.  These conclusions are based on the assumption that the merger remnant collapsed quickly into a black hole (motivated, e.g., by the lack of extra energy injection from a long-lived remnant), on the inferred mass of the ejecta combined with the inferred total mass of the two pre-merger neutron stars, and on relations between the maximum mass of a nonrotating neutron star and the maximum mass of a differentially rotating neutron star such as was (at least temporarily) produced by the merger.  All of these assumptions have some uncertainty to them; for example, if the remnant did not produce a black hole and also did not produce a strong poloidal magnetic field, there is no guarantee that there would be observable extra energy injection.  This, plus the complexity of the modeling, suggests caution in incorporating maximum mass upper limits in EOS constraints.

Another possible constraint on the equation of state comes from oscillations of the remnant of a merger of two neutron stars, before the remnant collapses into a black hole.  Current gravitational wave detectors are not sensitive enough at the high expected frequencies ($>$1~kHz) to see these signals, but it has been proposed that such oscillations might be seen in short gamma-ray bursts \cite{2019ApJ...884L..16C}.  Indeed, two short bursts observed with the Burst and Transient Source Experiment on the Compton Gamma-ray Observatory (BATSE) have two strong quasi-periodic oscillations (QPOs) each \cite{2023Natur.613..253C}.  Interpreting these oscillations as being from the post-merger remnant, \cite{2025ApJ...983...88G} found a 68\% credible range of $12.48^{+0.41}_{-0.40}$~km for the radius of a $1.4~M_\odot$ neutron star, which is compatible with other constraints and is one of the tightest bounds yet found.  Recently, a single strong QPO candidate has been reported from a third BATSE burst \cite{2026ApJ...998..289C}, which could be consistent with the post-merger oscillation interpretation.

\subsection{The future: moment of inertia}
\label{sec:MomIntertia}

A development to track is the ongoing improvement in the precision of the measurement of the moment of inertia of the faster-spinning star in the double pulsar binary PSR~J0737$-$3039.  This binary, which was discovered in 2003 \cite{2003Natur.426..531B}, is unique in that both neutron stars, rather than just one, appeared to us as pulsars (until 2008, when the beam from the slower-spinning of the two precessed out of our line of sight \cite{2010ApJ...721.1193P}).  The masses are known to better than 0.1\%, and the faster-rotating of the two neutron stars (PSR~J0737$-$3039A, with a rotation period of 23 milliseconds) drags spacetime sufficiently to produce a measurable extra precession of the pericenter of the orbit.  The challenge is that the precession is overwhelmingly dominated by the lowest-order pericenter precession (the same one that gives Mercury an extra $43"$ of precession per century in its orbit), which is dominated by the total mass of the binary.  Hence that contribution has to be known precisely enough that the extra precession due to frame-dragging can be identified clearly.  It is anticipated that the extra precession, and thus the moment of inertia, will be known to $\sim 10-11$\% by the year 2030 \cite{2020MNRAS.497.3118H,2024Univ...10..160H}.  Given that the moment of inertia scales as $\sim MR^2$ and the mass is known very well, this is roughly equivalent to a $\sim 5-6$\% measurement of the radius, although in detail there is a dependence on the structure of the star.

\section{NEUTRON STAR RADIUS MEASUREMENTS BEFORE NICER}
\label{sec:NSradpreNICER}

\begin{figure}[htbp]
\vskip-2.0truein
\includegraphics[width=15.0cm]{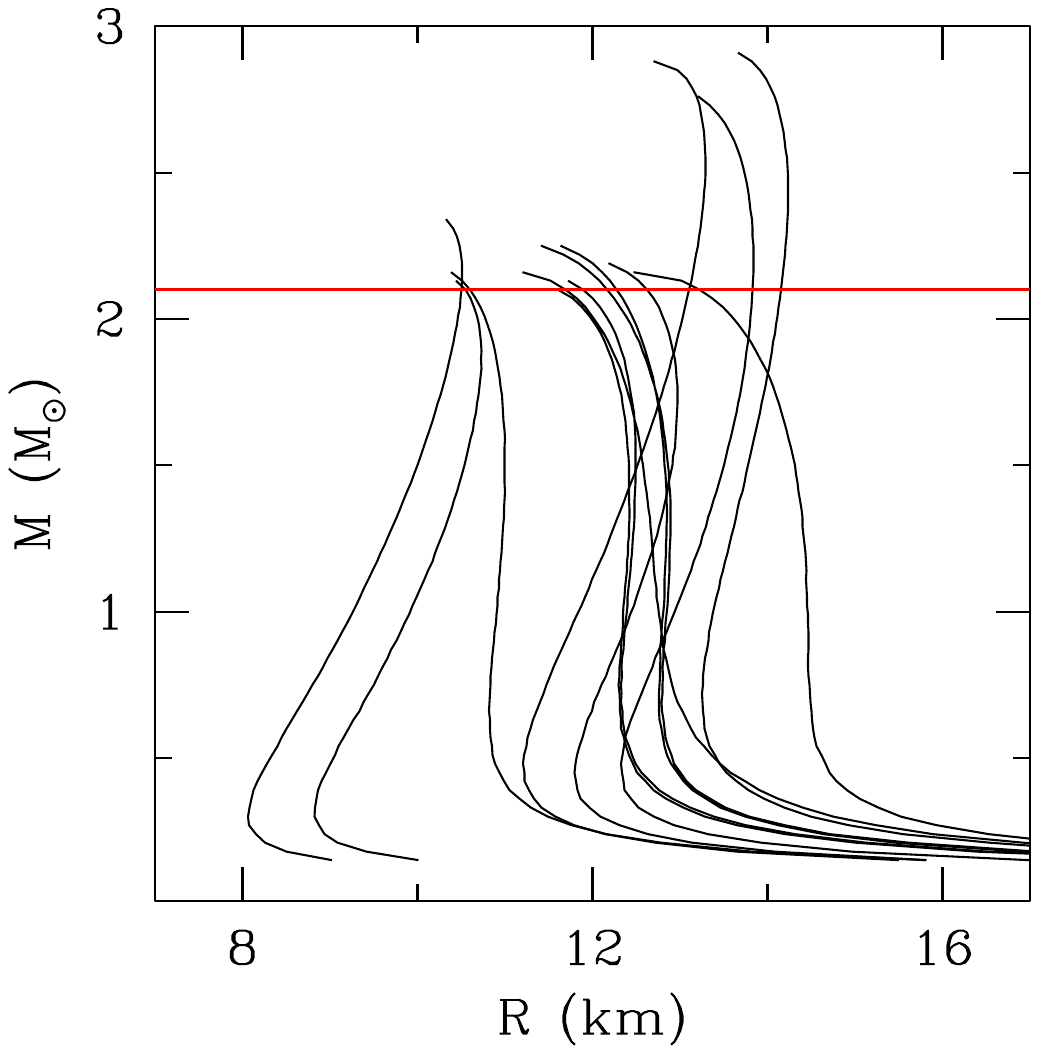}
\vskip-1.7truein
\caption{Mass-radius curves of different EOS constructed using the Gaussian processes approach detailed in \cite{2021ApJ...918L..28M}. The horizontal red bar is at $2.1~M_\odot$, which is the approximate mass of PSR~J0740$+$6620, the neutron star with the highest precisely measured mass.  All of the EOS displayed here have a maximum mass at least that large.  Nonetheless, the range of radii is large ($\Delta R\gtorder 4$~km at both $M=2.1~M_\odot$ and at a canonical mass $M=1.4~M_\odot$).  This highlights the importance of precise and accurate radius measurements.}
\label{fig:radrange}
\end{figure}

\textbf{Figure~\ref{fig:radrange}} gives an indication of why precise and accurate radius measurements are important.  This figure shows mass-radius curves selected from a set of EOSs which all have maximum masses of at least $2.1~M_\odot$ (see Section~\ref{sec:EOSconstraints} for how these EOSs are constructed).  Thus all of them can accommodate the highest neutron star masses yet observed.  However, we note that the radii at, for example, $1.4~M_\odot$ or $2.1~M_\odot$ vary widely among this set.  Thus radius measurements are essential to cull the set of viable EOS.

Radius measurements, however, are quite challenging.  We can get a sense of the difficulties by considering how we would measure the radius of an ordinary star.  Most stars cannot be angularly resolved with current telescopes, so we cannot use the simplest geometric measurement (angular radius times distance equals physical radius).  However, stars similar to the Sun emit nearly uniformly and isotropically and have spectra close to blackbodies, so we can obtain a reasonable radius estimate by measuring the distance $d$ (using, e.g., parallax) and flux $F$ to get the luminosity $L=4\pi d^2F$, and can measure the spectrum to find the temperature $T$, from which the radius $R$ follows from the blackbody formula $L=4\pi R^2\sigma_{\rm SB}T^4$, where $\sigma_{\rm SB}$ is the Stefan-Boltzmann constant.  Checks of the radii thus obtained with stars where we know the radius by other means (e.g., the Sun) give good agreement.

When the blackbody-fitting approach is applied to neutron stars, impossibly small radii are obtained.  This is because the atmospheres of non-accreting neutron stars are dominated by scattering, which reduces the efficiency of emission.  Even when accurate model spectra are used, fits to the X-ray data from cooling neutron stars are typically fit equally well with pure hydrogen or pure helium atmospheres (e.g., \cite{2012MNRAS.423.1556S}; the surface gravity of neutron stars is large enough that the lightest nucleus present in abundance is expected to float to the top).  However, under the assumption that the whole surface emits uniformly, hydrogen and helium fits often yield dramatically different best-fit radii.  For example, at 68\% credibility \cite{2022ApJ...941...76V} find that the quiescent low-mass X-ray binary in the globular cluster M28 has a $9.2-11.5$~km radius assuming that its atmosphere is pure unmagnetized hydrogen, and a $13.0-17.5$~km radius assuming that its atmosphere is pure unmagnetized helium.

Even if we know the atmospheric composition, the assumption of uniform emission is suspect.  For example, \cite{2010ApJ...722...33S} found that this assumption, combined with other standard treatments, is inconsistent with the data on the bursting source 4U~1820$-$30.  \cite{2014MNRAS.445.4218K} found that this inconsistency extends to a large fraction of the bursting sources analyzed using the assumption of uniform emission.  Because hydrogen and helium are both fully ionized at the typical temperatures of neutron star atmospheres, the spectrum is a continuum, with no lines or edges, which means that at the typical data quality even a wide range of temperatures on the surface cannot be distinguished from a single temperature, even though the radius inferred using a single temperature can be strongly biased \cite{2019Univ....5..100M}.  If uniform emission is nonetheless assumed \cite{2016ApJ...820...28O}, the inferred radii are small (typically $<11$~km).  If the uniform-emission assumption is relaxed and replaced with an extra parameter indicating the fraction of the surface that emits, then the inferred radii increase to $\sim 12-13$~km, at least for the neutron star 4U~1702$-$429 \cite{2017A&A...608A..31N}.  

Overall, the radii obtained using time-averaged fluxes and spectra are subject to potentially large systematic errors that are difficult to quantify.  Data of a different character are needed, and this was the motivation behind NASA's NICER mission.

\section{THE NICER MISSION AND ITS DATA}
\label{sec:NICER}

The NICER mission was launched on 3 June 2017 for deployment on the International Space Station.  Observations with NICER have contributed to a wide variety of studies of compact objects and X-ray emission, but our focus here is on its primary mission, which is to facilitate the measurement of neutron star masses and radii using precise pulse waveforms.

The key advance of NICER compared with previous X-ray missions is the combination of $<100$ nanosecond time-tagging of each X-ray photon (much shorter than the $>1$ millisecond rotation periods of neutron stars), good soft X-ray sensitivity (effective area of $\sim$few hundred cm$^2$ to $\sim 2000$~cm$^2$ in the 0.3~keV to 3~keV band), and few-percent spectral resolution.  This is well-matched to the characteristics of the non-accreting neutron stars that are NICER's primary targets (accreting neutron stars are much brighter in X-rays, but complexities of the accretion introduce potentially major systematic errors).  The NICER targets are observed as pulsars in radio waves, which means that precise radio timing allows each X-ray photon to be assigned a rotational phase.  Therefore, over years of real time, millions of seconds of non-contiguous observations can be folded into a single pulse waveform, which gives the X-ray spectrum as a function of rotational phase.  This is the product that is analyzed using the spot models that we now discuss.

The basic idea is that the production of the coherent radio waves that we see as radio pulsations involves the generation of cascades of high Lorentz factor electrons and positrons ($\gamma\sim 10^7$ is expected for the pulsars observed with NICER, which rotate hundreds of times per second \cite{2019ApJ...887L..25B}).  The electrons and positrons hit the surface, where they release energy that then comes out as thermal X-rays.  The emitted X-rays are concentrated in regions determined by the structure of the stellar magnetic field.  Colloquially we call these regions ``spots".  Stellar rotation, with the spots fixed to the surface, then produces a rotational dependence to the spectrum.  Analysis of these data provides constraints on the neutron star mass and radius as well as other quantities associated with the spots themselves.

A simplified picture helps us understand how we might get mass and radius constraints from the pulse waveform.  Suppose that we have a small spot on the rotational equator and we, the observer, are at a great distance but also in the rotational equatorial plane.  The star has a known rotational frequency $\Omega$ from radio and X-ray observations.  If we begin by imagining a star whose compactness $GM/(Rc^2)$ is small enough to ignore, then we can see the spot for half of a rotational cycle.  If the rotation speed is much less than the speed of light then the observed flux will peak when the spot is almost directly under us, because its projected area will then be maximal.  This is also to say that the pulse waveform will be nearly symmetric in its rise and fall.  However, if the rotational speed $v$ is not much less than $c$, then beaming competes with projected area, and the observed flux will peak at an angle ${\cal O}(v/c)$ before the spot is directly below us.  Hence the waveform will appear asymmetric, and in this circumstance we could infer $v$.  Doppler shifts of the spectrum will also inform us about $v$.  Knowledge of $v=\Omega R$ and $\Omega$ would then give us a measure of $R$.

If now we take into account that for a neutron star the compactness cannot be neglected, the next effect is that light bending will allow us to see the spot for more than half of a rotational cycle.  The fraction of the cycle during which we see the spot is a measure of the compactness, which means that in this simplified scenario the fraction of the cycle where the spot is visible plus the asymmetry of the pulse waveform gives us both the mass and the radius.

Of course, real observations will not be this simple: the spots are not infinitesimal, our line of sight and the colatitude of the spot must be determined from the data, there are background counts from high-energy particles and sources in the field of view of NICER, and so on.  Moreover, the shapes and temperature distributions of the spots are likely to be highly complex.  Fortunately, work prior to the launch of NICER, which has been elaborated and confirmed after the launch, suggests that as long as the spots are small (which in practice means angular sizes well under a radian), the details are of minimal importance \cite{2009ApJ...705L..36L,2009ApJ...706..417L,2015ApJ...811..144B}.  Even more encouragingly, investigations to date have found that if NICER-like data are fit well by a spot model (in the sense of, e.g., $\chi^2$), then the radius posterior is not significantly biased \cite{2013ApJ...776...19L,2015ApJ...808...31M}.  Of course, if the data are not fit well by a given model, then the corresponding radius inferences cannot be trusted.

With this in mind, it is possible to use simplified models of the spots.  Here we focus on the approach of our group (e.g., \cite{2019ApJ...887L..24M,2021ApJ...918L..28M,2024ApJ...974..295D}); for different choices, see \cite{2019ApJ...887L..21R,2021ApJ...918L..27R}.  

We model our spots as being either circular or ovate, and as having uniform effective temperature throughout their area.  The number of spots is specified in advance of a particular analysis; in practice we have not needed to use more than three spots, and none of the pulse waveform data sets that we have used from NICER can be fit well with only one spot.  The spots can have any size.  We number the spots 1, 2, 3, etc., and if there is an overlap between two or more spots then the pixels in the overlap region emit with the effective temperature of the lowest-numbered spot.  In addition to the parameters that describe the spots (center colatitude and longitude, angular radius, and effective temperature for circular spots; those four parameters plus a stretching factor and orientation angle for the ovate spots), each model specifies the inverse compactness $c^2R/(GM)$, the mass $M$, the observer inclination, the neutral hydrogen column density (a measure of the amount of interstellar absorption from the pulsar to the observer), and the distance to the pulsar.  In addition to these primary parameters, there are nuisance parameters related to unmodulated emission (``background") in each of the NICER energy channels.  Depending on the analysis, additional parameters could be included which, for example, take into account uncertainty in the calibration of NICER or cross-calibration with other instruments such as those on the X-ray Multi-Mirror (XMM-Newton) satellite.  

The analyses involve many parameters.  For example, even a two-circle analysis has 12 primary parameters, and in practice the likelihood surface can be multipeaked.  Thus, to obtain a thorough and representative posterior on the parameters, sophisticated samplers are needed, typically requiring $\sim 10^8$ samples.  We discuss these and other potential systematic errors in the next section.

\section{POTENTIAL SYSTEMATIC ERRORS IN NICER ANALYSES}
\label{sec:systematic}

As we indicated in Section~\ref{sec:NSradpreNICER}, neutron star radius estimates prior to NICER were susceptible to systematic errors that could be much larger than the formal statistical uncertainty in the fits.  The major advantage of radius estimates from NICER data is that, as far as is currently known from studies with synthetic data, if the best fit is statistically good then, assuming thorough statistical sampling, the radius posterior is not very biased (in practice, the radius used to generate the synthetic data is not systematically displaced by more than $\sim 1\sigma$ from the median of the posterior).  In this section we discuss several potential systematic errors in analysis of NICER data and present our current understanding about their impact.

\subsection{Spot shape and temperature distribution}
\label{sec:shape}

As discussed in Section~\ref{sec:NICER}, the spots on neutron stars which emit the X-rays observed with NICER are produced by the deposition of energy from high-energy electrons and positrons produced as part of the process of coherent radio emission in these pulsars.  The pattern of energy deposition on the surface is complicated and depends on the multipolar distribution of the stellar magnetic field, among other things.  The real shape is therefore neither circular, nor ovate, nor any other simple shape, and the energy deposition distribution and thus the effective temperature distribution is not uniform.  In contrast, our assumption in the fits is that the effective temperature is constant within the boundaries of spots with simple shapes.  

This sounds like a potentially troublesome mismatch.  However, \cite{2009ApJ...705L..36L,2009ApJ...706..417L,2015ApJ...811..144B} found that as long as the spots are small (e.g., dimension well under $\sim 1$~radian), the details of the spot shape and/or temperature distribution have only a small effect on the radius inference.  The actual spot shape could be a circle, oval, arc, or disconnected dots.  The temperature could be uniform or broadly distributed.  A fit using a simplified spot could get the angular radius or shape of the spot wrong, but the inferred radius will not be significantly biased.  The basic reason, as elucidated in \cite{2009ApJ...705L..36L,2009ApJ...706..417L}, is that for thermal emission from the surface the beaming pattern from a pixel on the surface is very broad, and then gravitational light deflection blurs the pattern further.  Thus small-scale angular details are lost.  Put another way, if one decomposes the pulse waveform into Fourier components, the breadth of the beaming combined with gravitational light deflection strongly reduces the small-angle components.

The robustness against spectral errors can be explained by the fact that the spectrum is a continuum, without lines or edges (this follows from the expected hydrogen or helium composition of the atmosphere and the lack of spectral features in the NICER energy band of $E>0.3$~keV).  As an example of how forgiving continuum fitting can be, \cite{2019Univ....5..100M} performed a fit with a single-temperature blackbody of synthetic data constructed with full-surface emission from a neutron star that had a factor of 10 change in temperature from equator to pole.  Nonetheless, the single-temperature model, with $\sim 3\times 10^4$ counts in the synthetic data set, fit the data well.  The inferred area of a spot with the wrong assumed temperature distribution could be incorrect, but as long as the spot is small these issues will not compromise the radius inference.

If instead the spot angular size is large in some dimensions, a fit with the wrong spot model could be poor.  For example, the best fit to the NICER data from PSR~J0030$+$0451 has one highly elongated spot with a long dimension more than a radian \cite{2019ApJ...887L..24M,2019ApJ...887L..21R}.  This is large enough that a circular model does not fit well.  In such a case, a fit with circular spots would raise red flags because of the poor fit, and it would then be necessary to explore models with greater complexity.

\subsection{Surface beaming pattern and composition}
\label{sec:beaming}

As mentioned earlier, the thermal X-ray emission from the spots is believed to come from energy deposition due to the impact of high Lorentz factor electrons and positrons on the surface, where those electrons and positrons are produced as part of the coherent radio emission.  NICER analyses are conducted under the ``deep heating" assumption --- that is, for a given rate of energy deposition, the intensity of emergent radiation is calculated as if the energy were generated at infinite depth.  As discussed by \cite{2019ApJ...872..162B,2020A&A...641A..15S}, this is a good approximation if the average Lorentz factor of the electrons and positrons exceeds $\gamma\sim 100$, and because the average Lorentz factor for millisecond pulsars is likely to be $\sim 10^7$ \cite{2005ApJ...622..531H}, this is probably a good assumption.  But what if there is a substantial lower-energy tail?  Then \cite{2019ApJ...872..162B,2020A&A...641A..15S} show that the energy is deposited in a shallower layer.  The result is that the emergent intensity is not beamed as much along the surface normal as it would be for deep heating.  Another way that the beaming pattern could be changed is if the atmosphere were composed of helium rather than hydrogen (see Figure~1 from \cite{2021ApJ...914L..15B}).

However, \cite{2023ApJ...956..138S} showed that (1)~the choice of atmosphere makes a negligible difference to the radius inferred from NICER and XMM-Newton data on the key high-mass pulsar PSR~J0740$+$6620 (see also the discussions in \cite{2021ApJ...918L..28M,2024ApJ...974..295D}), (2)~although the inferred radius for PSR~J0030$+$0451 \textit{is} changed when the atmospheric composition is assumed to be helium, the evidence favors hydrogen (see also \cite{2019ApJ...887L..24M}), (3)~realistic beaming deviations assuming that (improbably) the typical Lorentz factor is $\ltorder 100$ have only a small effect on the inferred radius.  Thus in practice beaming deviations are not likely to introduce substantial systematics in the inference of neutron star radii from NICER data, although given the importance of this question other investigations are ongoing (I. Holt, in preparation).

\subsection{Magnetic fields}
\label{sec:magfields}

A sufficiently strong magnetic field can alter the beaming pattern or spectrum from a point on a neutron star surface in two ways.  First, it can change the structure of atoms in the atmosphere and thus the opacities.  Second, it can change the process of scattering off of free electrons.  The most relevant comparison for the purposes of NICER analyses is the photon energy versus the electron cyclotron energy.  Because the lowest-energy channel we analyze corresponds to a photon energy of $\approx 0.3$~keV, the magnetic field for that cyclotron energy would be $B\approx 3\times 10^{10}$~G.  For the rapidly rotating stars that are the focus of NICER observations, the magnetic dipole moments inferred from the gradual slowdown of rotation are $\sim {\rm few}\times 10^8$~G (see, e.g., \cite{2019ApJ...887L..24M,2021ApJ...918L..28M}) and thus the magnetic field probably does not play a major role.  

This is, however, not rigorously established.  The sometimes complex spot patterns that best fit NICER data are evidence that the magnetic field is not a simple centered dipole.  In principle, if there are strong multipolar components to the field then the surface field strength could be larger and the magnetic effects correspondingly greater.  A full assessment of the effect of magnetic fields would require models that we do not have: full opacity tables and model atmospheres for a wide range of magnetic fields and angles to the surface normal.  This would require an effort many times greater than what has already been invested in nonmagnetic atmospheres.  In the meantime, the excellent fits to NICER data using nonmagnetic atmospheres serve as crude evidence that magnetic effects are not dominant.

\subsection{Other components: background, nonthermal emission, information from XMM-Newton}
\label{sec:other}

NICER data contain contributions from the spots, but also from other components including background and, possibly, modulated nonthermal emission from the star.  If a component is not included then its absence could lead to systematic errors in the inferred radius.  For example, PSR~J0437$-$4715 has pulsations detected with NuSTAR \cite{2016MNRAS.463.2612G} at several keV, which is well above the energy that could be provided from a thermal component on the surface.  If this component is not included in the model, it risks biasing the fitted radius.  

For the background, our assumption is that any emission which does not vary in a way commensurate with the rotation frequency of a pulsar contributes equally to all rotational phases, given that the observation time is many orders of magnitude longer than the rotational period.  Our most conservative approach, making the fewest a priori assumptions, is to assume nothing about the background and, instead, allow any nonnegative amount of background in each NICER energy channel.  However, there are some fits to what are at least components to the background, e.g., the 3C50 model of \cite{2022AJ....163..130R}, which models the part of the background provided by the small fraction of high-energy charged particles that evade particle vetos in NICER.  Inclusion of such a background will not bias radius inference as long as the background model is reliable.  Of course, if the information included is not reliable then bias is a risk.  For example, if the assumption in the analysis is that the particle background is the only background, but in reality there is other background, then this will incorrectly push the spot model to make up the extra counts.

A similar situation exists with the use of data from other instruments.  In particular, data from XMM-Newton have been used for several NICER pulsars.  XMM-Newton has a much lower background than NICER, which means that a plausible lowest-order assumption is that all of the counts seen in XMM-Newton observations come from the pulsars themselves.  This assumption is not quite accurate, because there are also counts from the XMM-Newton field of view from angularly nearby sources including distant unresolved sources.  Observations can provide estimates of the flux from these background sources, but there could be additional contributions that are not included (e.g., from an intrabinary shock in a pulsar binary).  Fortunately, recent work \cite{2025arXiv251116759H} suggests that even a factor of 5 mismodeling of this background emission from XMM-Newton does not change the inferred radius by more than roughly one standard deviation.

The final issue to discuss in this subsection is the calibration of NICER and, when applicable, of other instruments such as those on XMM-Newton.  Absolute calibration of the response (i.e., the effective area with which a photon of a given energy is read as a count in a given energy channel) of X-ray detectors in space is a highly challenging endeavor.  This is because the true X-ray spectrum of astronomical sources is not known, and broadband artificial calibration sources do not exist in the X-rays.  However, it is believed based on extensive cross-calibration with other instruments that NICER's calibration is no more than a few percent off \cite{2021ApJS..255....7R}.  It is also believed, based on observations of sources thought to have power-law spectra, that if NICER's calibration is in error the first-order correction is a constant multiplier of the effective area, independent of energy (see the discussion in Section~3.7 of \cite{2021ApJ...918L..28M}).  Thus a single extra parameter can accommodate the most important calibration deviations.

\subsection{Statistical sampling and incompleteness}
\label{sec:sampling}

As discussed earlier, even the simplest models applied to NICER data have a substantial number of parameters and potentially multimodel likelihood surfaces.  Thus for the posterior to be correct, it is necessary to have sophisticated and reliable statistical samplers.  This is more easily said than done.  As a thought experiment demonstrates, a sufficiently pathological likelihood will defeat any sampler: imagine, for example, a likelihood which is enormous in such a small range of parameter space that it cannot be found except by luck.  Even without such pathologies, near-degeneracies between parameters and fits with similar quality in different parts of parameter space need to be sampled thoroughly.

Over the last decade plus, several Bayesian sampling codes have emerged.  In a broad sense, one category (called nested samplers \cite{2004AIPC..735..395S}; a current example is {\tt MultiNest}, see \cite{2009MNRAS.398.1601F}) is primarily designed for model comparison, whereas another category (an example of which is Markov chain Monte Carlo samplers, e.g., the code {\tt emcee}, see \cite{2013PASP..125..306F}) is primarily designed for parameter estimation within a given model.  Each code has precision parameters that can be set, and to the degree that it has been tested it appears that when the precision is high enough the codes agree with each other.  However, in practice the precision parameter values often used for {\tt MultiNest} are insufficient: they underestimate the breadth of the posterior distribution in radius, sometimes by tens of percent \cite{2019ApJ...887L..24M, 2021ApJ...918L..28M, 2021AJ....162..237I, 2023MNRAS.521.1184L, 2024OJAp....7E..79D, 2024ApJ...974..295D,2025PhRvD.112b3008H}.  Thus, a question to ask when using a NICER-based radius result, especially for results obtained with {\tt MultiNest}, is the following: Has there been a convergence study with different levels of precision in all tunable parameters in the statistical sampling?  If not, then the quoted uncertainties might be underestimated.

\section{RESULTS FROM ANALYSES OF NICER DATA}
\label{sec:results}

In this section we summarize briefly the results from each of the five pulsars with radii reported from analyses of NICER data, and comment on potential caveats for each analysis.

\begin{figure}[htbp]
\vskip-1.7truein
\includegraphics[width=15.0cm]{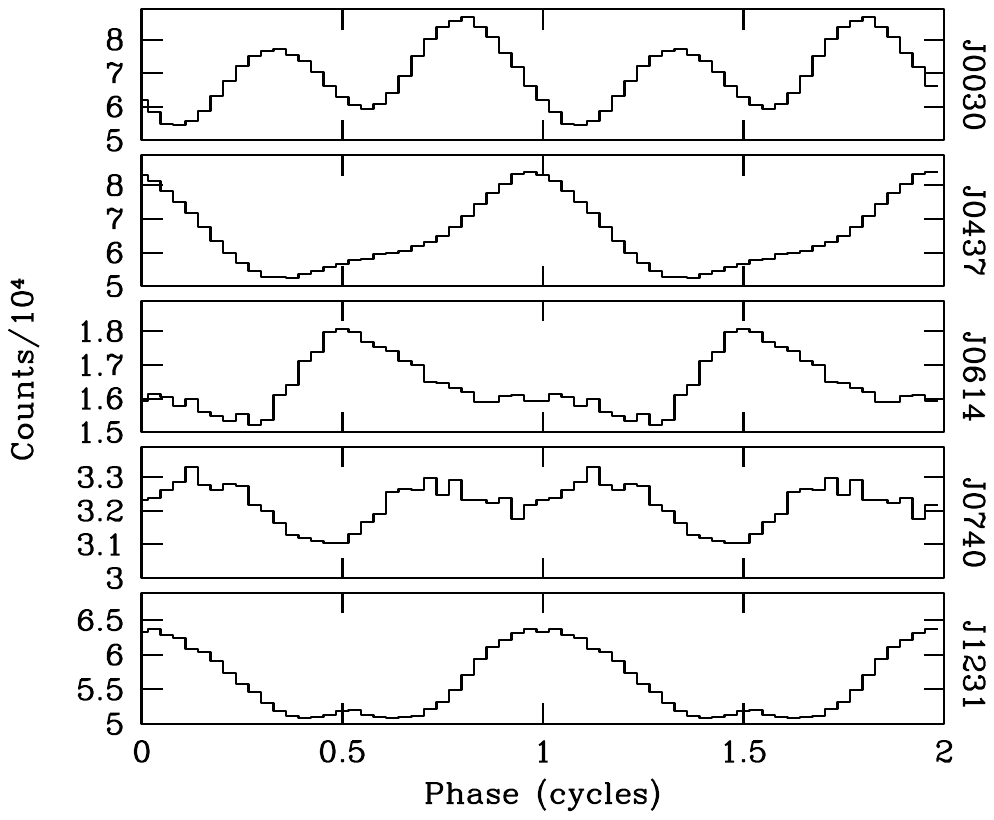}
\vskip-2.7truein
\caption{Bolometric waveforms for each of the five pulsars with reported NICER analyses.  The horizontal axis gives the rotational phase (repeated over two full cycles to show patterns) and the vertical axis gives the number of counts in each of the 32 rotational phases in units of $10^4$ counts.  Each panel is labeled on the right with the name of the pulsar.  Note that the number of counts depends on which set of observations are included and which NICER energy channels are summed (e.g., for PSR~J0740$+$6620 the counts are summed over NICER PI channels 30 through 123 inclusive, for PSR~J1231$-$1411 the range is channels 30 through 149 inclusive, for PSR~J0614$-$3329 the range is channels 30 through 200 inclusive, and for PSR~J0030$+$0451 and PSR~J0740$+$6620 the range is channels 30 through 299 inclusive), but the shapes are characteristic of each pulsar.  We see from this figure that (1)~the shapes of the waveforms differ substantially from pulsar to pulsar, (2)~in detail, no pulsar fits the expectations of the symmetries of antipodal and identical spots (which thus means that more complex models are required), and (3)~the ratio between the modulated component and the DC component has a large range between the pulsars (e.g., PSR~J0030$+$0451 has $\sim 40$\% as many modulated counts as DC counts, whereas PSR~J0740$+$6620 has $\sim 10$\% as many modulated counts as DC counts), which means that background treatments are more important for some pulsars than for others.  Data from https://heasarc.gsfc.nasa.gov.}
\label{fig:bolo}
\end{figure}

\subsection{PSR~J0030$+$0451}
\label{sec:J0030}

The first of the NICER pulsar analyses was for PSR~J0030$+$0451, which has a rotation frequency of 205.53~Hz and is an isolated pulsar, which means that it is not possible to use binary orbital modulation to obtain independent information about the mass or orbital inclination.  \cite{2019ApJ...887L..24M} found a mass of $M=1.44^{+0.15}_{-0.14}~M_\odot$ and a radius of $R=13.02^{+1.24}_{-1.06}$~km, both at 68\% credibility, and using the same data the independent analysis of \cite{2019ApJ...887L..21R} found $M=1.34^{+0.15}_{-0.16}~M_\odot$ and $R=12.71^{+1.14}_{-1.19}$~km, also at 68\% credibility.   \cite{2019ApJ...887L..24M} and \cite{2019ApJ...887L..21R} used different spot models and somewhat different priors from each other.  Nonetheless, not only do their inferred masses and radii largely overlap, but their inferred emission regions on the star are extremely similar, bolstering confidence in both analyses.  Both papers used only the NICER data for this pulsar, and found that to be consistent with the XMM-Newton data, additional XMM background was likely to be necessary.

More recently, \cite{2026arXiv260223743K} analyzed an updated NICER data set (with $\sim 50$\% more counts than in the previous analysis), including a joint analysis with the XMM-Newton data, and found $M=1.43^{+0.20}_{-0.17}~M_\odot$ and $R=12.68^{+1.31}_{-1.04}$~km.  Importantly, \cite{2026arXiv260223743K}, which used MultiNest to sample the posterior, performed runs at different MultiNest resolutions and found that the runs gave similar answers; a final step which would be useful would be to put the MultiNest posterior into an MCMC sampler such as \texttt{emcee} to determine whether the posterior broadens.  This, however, is a promising result.  One of the assumptions in the \cite{2026arXiv260223743K} analysis is that there is no unknown background in the XMM-Newton data; this assumption may or may not be correct, and it might have affected some earlier joint analyses \cite{2024ApJ...961...62V}.

\subsection{PSR~J0740$+$6620}
\label{sec:J0740}

PSR~J0740$+$6620 has a rotational frequency of 346.53~Hz and is the neutron star with the highest precisely measured mass ($2.08\pm 0.07~M_\odot$; see Section~\ref{sec:NSmass}).  It is thus particularly important for EOS constraints because the central density of PSR~J0740$+$6620 is close to the highest density realized in neutron stars.  Hence, forms of matter (e.g., dominated by hyperons or involving free quarks) might exist in PSR~J0740$+$6620 that do not exist in $\sim 1.4~M_\odot$ neutron stars such as PSR~J0030$+$0451, PSR~J0437$-$4715, and the lower-mass neutron stars whose tidal deformability might be detected from their gravitational waveforms (see Section~\ref{sec:GW170817}).  

PSR~J0740$+$6620 is also a very faint source.  Its total NICER count rate is $\sim 0.4$ counts per second, but only $\approx 10$\% of that is from the spots, with the remainder being supplied by background (see Figure 5 of \cite{2024ApJ...974..295D}).  Thus additional information must be included for the radius to be inferred with useful precision.  This information comes from radio observations (which supply the mass, observer inclination, and distance) as well as XMM-Newton observations (which, because of the much lower background from XMM-Newton compared with NICER, provide a good constraint on the total number of counts from the star).  

After the initial analyses in 2021 \cite{2021ApJ...918L..28M,2021ApJ...918L..27R}, refined measurements were published in 2024 \cite{2024ApJ...974..295D,2024ApJ...974..294S}.  Dittmann et al. \cite{2024ApJ...974..295D} reported a 68\% credible radius range of $R=12.92^{+2.09}_{-1.13}$~km, whereas Salmi et al. \cite{2024ApJ...974..294S} found a 68\% credible radius range of $R=12.49^{+1.28}_{-0.88}$~km.  Most of the difference between these inferences is because \cite{2024ApJ...974..294S} required that the radius be less than 16~km, whereas \cite{2024ApJ...974..295D} allowed the radius to be anything that fit the data.  When \cite{2024ApJ...974..295D} also imposed a 16~km maximum, they found that at 68\% credibility the radius is $R=12.76^{+1.49}_{-1.02}$~km.  Moreover, when \cite{2024ApJ...974..294S} used a higher-precision setting of {\tt MultiNest} than they featured in their headline results, they found $R=12.55^{+1.37}_{-0.92}$~km, which is closer to the results of \cite{2024ApJ...974..295D} in both median and breadth; this follows the pattern that for analyses such as these, {\tt MultiNest} underestimates the median and standard deviation of the radius unless exacting precision settings are used.

\subsection{PSR~J0437$-$4715}
\label{sec:J0437}

PSR~J0437$-$4715 has a rotational frequency of 173.68~Hz and is the non-accreting millisecond pulsar with the highest NICER count rate, with roughly 1 count per second inferred to come from the spots.  It is also an unusually complex source; \cite{2016MNRAS.463.2612G} found in NuSTAR data that it has a modulated nonthermal tail.  Thus, in addition to the modulation provided by thermal emission from the spots, there is another component that contributes to modulation in the NICER data.  The first published analysis of NICER data on PSR~J0437$-$4715, by Choudhury et al. \cite{2024ApJ...971L..20C}, acknowledged the modulated nonthermal component but included only thermal modulation in their analysis.  Their headline result is a radius, at 68\% credibility, of $R=11.36^{+0.95}_{-0.63}$~km.  However, the bolometric fit to the data is not good, which suggests that the modulated nonthermal component might be necessary for a good fit.  Moreover, {\tt MultiNest} was used as the main parameter estimation tool, which in analogy with similar circumstances suggests that a fully converged analysis might result in larger uncertainties in the radius.  

More recently, Miller et al. \cite{2026ApJ..1000L..48M} analyzed the NICER data on PSR~J0437$-$4715 which included a modulated power law, to simulate the modulated nonthermal component seen in the NuSTAR data.  They did find an acceptable bolometric fit to the data, which suggests that this model might be closer to a realistic description of the system.  Their results used the intelligent MCMC-based sampler \texttt{pocoMC} \cite{2022MNRAS.516.1644K}, and converged between different resolutions.  They found a radius $R=14.12^{+1.36}_{-2.24}$~km.

\subsection{PSR~J1231$-$1411}
\label{sec:J1231}

As can be seen in \textbf{Figure~\ref{fig:bolo}}, the NICER data for the 271.45~Hz pulsar PSR~J1231$-$1411 show a weak interpulse feature.  This makes good fits challenging; \cite{2024ApJ...976...58S} found that the inferred radius depends strongly on the prior and is driven to unphysically small radii (below 8~km) if the radius is entirely free, to an improbably but not impossibly low mass ($M=1.04^{+0.05}_{-0.03}~M_\odot$) with an improbably precise radius ($R=12.6\pm 0.3$~km) when the radius is restricted by previous observations, and to a much larger but still unexpectedly precise radius ($R=13.5^{+0.3}_{-0.5}$~km) when the radius is required to be between 10~km and 14~km.  This is why the title of \cite{2024ApJ...976...58S} refers to PSR~J1231$-$1411 as ``A Complex Case" and why they note that better-fitting solutions will likely be necessary.

\subsection{PSR~J0614$-$3329}
\label{sec:J0614}

The most recent pulsar whose radius has been inferred from NICER data is the 317.59~Hz pulsar PSR~J0614$-$3329.  This is another low-flux source, but by combining XMM-Newton data with NICER data, \cite{2025arXiv250614883M} found a radius of $10.29^{+1.01}_{-0.86}$~km, to go along with the mass $M=1.44^{+0.06}_{-0.07}~M_\odot$ which was obtained using radio timing.  However, because the results were obtained using low-precision {\tt MultiNest} parameters, the converged posterior likely would be substantially broader than what has been reported.  This result also depends strongly on the XMM-Newton data, including the assumption that the XMM-Newton background is well-understood.  Nonetheless, if this small radius is confirmed it will contribute to our understanding of the EOS.

\section{CONNECTION BETWEEN NICER RESULTS AND NUCLEAR PHYSICS}
\label{sec:nuclear}

As described in Section~\ref{sec:NSprop}, the macroscopic properties of neutron stars are linked to microscopic nuclear properties through the EOS.  Thus in a broad sense we constrain the EOS and therefore the nuclear physics of cold, catalyzed, high-density matter in a Bayesian framework: starting from some set of candidate EOS, we compare their predictions with neutron star observations (e.g., the mass and radius from the TOV equation) and update the relative weights of those candidate EOS.  Here we discuss this procedure, starting with assumptions about the EOS below nuclear saturation density (and the resulting uncertainty) and then proceeding with various frameworks to describe the unknown high-density EOS.  

\subsection{Low-density EOS}
\label{sec:lowden}

The mass, radius, tidal deformability etc. of a neutron star depend on its entire structure, from its center to its surface.  It is commonly assumed that below nuclear saturation density the EOS is known fairly well because such densities can be accessed in laboratories.  A caveat is that in laboratories the equilibrium composition is fairly close to symmetric between neutrons and protons, whereas under the tremendous confining pressure in neutron stars protons can be outnumbered by 10:1 or more, depending on the density.  Uncertainties are mitigated by the small fraction of a neutron star's mass and radius that is in matter below saturation density; just a few percent of the mass, for example \cite{1983bhwd.book.....S}.  It has been estimated that the uncertainty in the equation of state of the solid crust of a neutron star (below $\sim$half of saturation density; \cite{2013ApJ...773...11H}) could lead to an uncertainty of up to $\sim 0.3$~km in the radius of a neutron star \cite{2016PhRvC..94c5804F,2020CQGra..37b5008G}.  This is small compared with the current statistical uncertainties in the radius from NICER observations, so the selection of the low-density EOS is not a limiting factor in the overall EOS inference.  Below half of saturation density we use the QHC19 EOS \cite{2019ApJ...885...42B}, but older EOS such as the SLy4 EOS of \cite{2001A&A...380..151D} and more updated EOS such as QHC21 \cite{2022ApJ...934...46K} give very similar answers.

\subsection{High-density EOS and possible constraints from nuclear physics}
\label{sec:highden}

At nuclear saturation density, some experimental data can be incorporated, such as the nuclear symmetry energy ($S=32\pm 2$~MeV; \cite{2012PhRvC..86a5803T}); its slope with density $L$ from experiments such as PREX-II \cite{2021PhRvL.126q2502A} and CREX \cite{2022PhRvL.129d2501A}; and the correlation between these and also with higher derivatives of the symmetry energy (e.g., \cite{2020PhRvC.102d5807L}).  At densities higher than saturation, frameworks such as chiral effective field theory (\cite{1990PhLB..251..288W,1991NuPhB.363....3W,1992PhLB..295..114W}; see \cite{2020PhRvL.125t2702D} for a recent review and \cite{2025PhRvL.135b2501C} for a recent reassessment of some specific three-nucleon forces in this context), can provide some insight but their formal uncertainties become large by $\sim 2\times$ saturation density, which is still well under the maximum reached in neutron stars.

At the other end of the density scale, equilibrium infinite-density matter is believed to consist of noninteracting quarks and gluons, which would thus behave as a fully relativistic free gas with EOS $p=\frac{1}{3}\epsilon$.  A question which has received recent attention is whether at lower densities perturbative quantum chromodynamics (pQCD) can be used to constrain neutron star matter.  The estimated uncertainty in the pressure of matter reaches tens of percent at $\approx 40\times$ saturation density (e.g., \cite{2018PhRvL.121t2701G}), and the largest densities reached in neutron star cores are currently estimated to be a few times saturation, so the gap is large.  Nonetheless, \cite{2023JHEP...06..002G,2023ApJ...955..100G,2024PhRvD.109i4030K} find that, using a limited set of EOS that bridge the density gap, considerations of causality ($dp/d\epsilon<c^2$) and stability ($dP/d\epsilon\geq 0$) can place interesting constraints on the EOS at densities accessible in neutron stars.  In contrast, \cite{2024EPJWC.29603002M,2024PhRvD.110l3009M} find, using a broader set of EOS, that pQCD considerations do not substantially constrain the EOS of neutron star cores.

\subsection{Current constraints on the EOS}
\label{sec:EOSconstraints}

Given a low-density EOS, we need to choose how to extend to higher densities so that we can use the TOV equation and other equations to predict maximum masses, radii, tidal deformabilities, and so on to compare with data, and then to update our relative weightings of the EOS that we employ.  Rather than using specifically tabulated EOS from the literature, it is more useful to select a framework by which we can generate a large number of EOS to evaluate.  There is, however, no unique way to do this.  For example, we could use a piecewise polytrope, in which in contiguous density intervals the pressure is $P\propto\rho^\Gamma$ with potentially different values of $\Gamma$ at different densities (see \cite{2009PhRvD..79l4032R} for an early example in this context; \cite{2016ApJ...831...44R} recommend five densities for such parameterizations).  Another approach is the spectral method of \cite{2010PhRvD..82j3011L,2018PhRvD..97l3019L}, in which the polytropic index $\Gamma$ changes continuously with density.  Yet another approach is to use Gaussian processes (introduced in this context by \cite{2020PhRvD.101l3007L}; see \cite{2023PhRvD.108d3013E,2024PhRvD.110l3009M,2024EPJWC.29603002M} for ways to incorporate phase transitions).  

Regardless of the framework, we follow the Bayesian approach of \cite{2020ApJ...888...12M} to incorporate astrophysical constraints including their uncertainties.  In this approach, the likelihood of a given data set given an EOS is computed in different ways depending on the nature of the data.  For example, as discussed in Section~\ref{sec:NSmass}, high neutron star masses rule against EOS with maximum masses that are too low (with proper accounting for uncertainties in the mass measurements; see \cite{2016EPJA...52...69A,2020ApJ...888...12M}).  Because the maximum mass is a property of the EOS, for a given candidate EOS the maximum mass can be computed and compared directly with the measured mass plus uncertainties to provide the likelihood.  In contrast, when comparing with data on the radius or tidal deformability of a star there is an extra (or ``nuisance") parameter, e.g., the central density.  Thus to compute the likelihood of such data for a given EOS we need to marginalize over the central density.  The prior on the central density is not known from first principles, but fortunately the choice (e.g., between a flat prior and a quadratic prior on the central density; see \cite{2021ApJ...918L..28M}) does not change the EOS posteriors substantially.

With these considerations in mind, our final choice is which data sets to incorporate in our analysis.  We choose (1)~a prior on the symmetry energy (energy per nucleon in pure neutron matter minus energy per nucleon in symmetric nuclear matter) at nuclear saturation density of $S=32\pm 2$~MeV \cite{2012PhRvC..86a5803T}, (2)~the high mass of the pulsar PSR~J1614$-$2230 (note that the higher-mass pulsar PSR~J0740$+$6620 is included under radius measurements), (3)~the tidal deformability from GW170817, (4)~our NICER-derived mass and radius for PSR~J0030$+$0451 \cite{2019ApJ...887L..24M}, and (5)~our NICER+XMM-derived radius for PSR~J0740$+$6620 \cite{2024ApJ...974..295D} combined with its radio-measured mass.  We have elected not to include chiral effective field theory constraints because we are focusing on the impact of astronomical measurements.  The additional astronomical constraints discussed elsewhere in this review would tighten the posteriors, but in our opinion there are enough questions about systematics and/or sampling thoroughness that inclusion of these constraints would risk the introduction of bias.

\begin{figure}[h]
\vskip-2.0truein
\includegraphics[width=18.0cm]{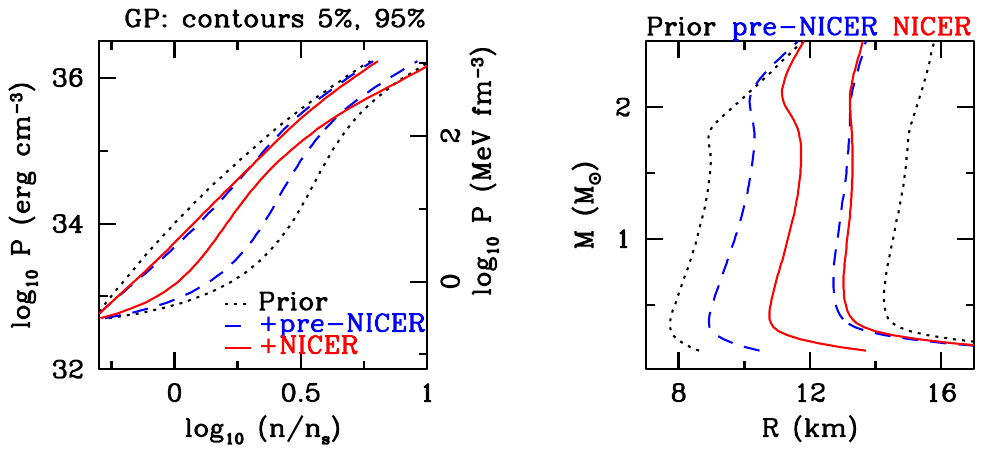}
\vskip-5.5truein
\caption{Left panel: pressure vs. density curves with different levels of constraints.  The horizontal axis gives the $\log_{10}$ of the number density $n$ of baryons in units of nuclear saturation density $n_s=0.16$~fm$^{-3}$.  The EOS set was constructed at high densities using Gaussian processes (GP; see \cite{2021ApJ...918L..28M} for more details).  For each line type the lower line gives the 5th percentile of the pressure at a given density and the upper line gives the 95th percentile of the pressure at a given density.  The dotted black lines are for our prior, the dashed blue lines are for our prior plus pre-NICER information (constraints from the symmetry energy at nuclear saturation density, from the high masses of a few pulsars, and from the tidal deformability constraints stemming from gravitational wave observations of GW170817), and the solid red lines are when we also add constraints from our NICER and XMM-Newton based measurements of the radii and masses of PSR~J0030$+$0451 and PSR~J0740$+$6620.  Right panel: mass versus radius constraints using the same sets of data as in the left panel.  For each line type the left line gives the 5th percentile of the radius at a given mass and the right line gives the 95th percentile of the radius at a given mass.  Above $M\sim 2~M_\odot$ the contours tend to slope up and to the right because the EOS that can reach high masses have high pressures at the relevant densities, which leads to larger radii.  We see that NICER measurements tighten the range of allowed EOS substantially, particularly at densities $\sim 1-3n_s$, and that NICER observations have dramatically tightened the allowed radii for neutron stars.  Left panel adapted from \cite{2026ApJ..1000L..48M} (CC BY 4.0). Right panel includes data from \cite{2021ApJ...918L..28M}.}
\label{fig:eosrad}
\end{figure}

The results of the analysis are displayed in \textbf{Figure~\ref{fig:eosrad}}.  In the left panel we show the 5th to 95th percentile in pressure as a function of baryon number density using the Gaussian process framework for our prior (black dotted lines), our prior plus pre-NICER results (blue dashed lines), and finally also including our NICER results (red solid lines). The results are similar for piecewise polytropes and spectral parameterizations of the EOS (see Figure~9 from \cite{2021ApJ...918L..28M} for a more comprehensive version).  From this panel it is clear that incorporation of the NICER and XMM-Newton radius measurements has tightened the EOS substantially, and in particular has hardened the EOS (i.e., has increased the pressure) from $\sim 1-3\times$ saturation density.  The effect of the radius measurements is seen even more clearly in the right hand panel of \textbf{Figure~\ref{fig:eosrad}} which shows the 5th to 95th percentile in radius as a function of mass, using the same line types as in the left hand panel.  The improvement using NICER data is evident.

\section{SUMMARY}
\label{sec:summary}

Radius measurements of neutron stars using NICER X-ray data have substantially enhanced our knowledge of the properties of cold catalyzed matter beyond nuclear saturation density.  For example, the uncertainty in the pressure at $\sim 2\times$ saturation density is roughly half what it was prior to NICER observations, as is the uncertainty in the radius at a canonical $M=1.4~M_\odot$ for neutron stars.  It also appears that the radii of $\sim 1.4~M_\odot$ stars such as PSR~J0030$+$0451 are similar to the radii of $\sim 2.1~M_\odot$ stars such as PSR~J0740$+$6620.  This and other clues suggest that at densities accessible to neutron stars the sound speed $c_s$ is greater than the conformal sound speed, $c_s^2>\frac{1}{3}c^2$, although the case is not yet airtight \cite{2024PhRvD.110l3009M}.  If further improvements in precision confirm this inequality then it indicates an interestingly nontrivial dependence of sound speed: given that at infinite density $c_s^2=\frac{1}{3}c^2$ and that perturbations at high but finite densities lead to $c_s^2<\frac{1}{3}c^2$ \cite{2023JHEP...06..002G}, this would imply that the sound speed reaches at least one local maximum, then at least one local minimum, as the density increases.  This could present clues to newly emerging degrees of freedom or interactions between quarks.

Additional work still needs to be done with the NICER data.  For example, all pulsars analyzed jointly with XMM-Newton data need to use the most updated calibrations.  Further studies also need to be performed to explore potential systematic errors.  However, NICER analyses have revolutionized our understanding of cold dense matter and are likely to be the primary source of information on the most massive neutron stars for many years to come. 

\section*{DISCLOSURE STATEMENT}

The author is not aware of any affiliations, memberships, funding, or financial holdings that might
be perceived as affecting the objectivity of this review.

\section*{ACKNOWLEDGMENTS}

M.C.M. was supported in part by NASA ADAP grant No. 80NSSC21K0649. Part of this work was performed at the Aspen Center for Physics, which is supported by U.S. National Science Foundation grant No. PHY-2210452. We also thank the Institute for Nuclear Theory at the University of Washington, which is supported in part by U.S. Department of Energy grant No. DE-FG02-00ER41132. Some of the resources used in this work were provided by the NASA High-End Computing (HEC) Program through the NASA Center for Climate Simulation (NCCS) at Goddard Space Flight Center, and also by supercomputing at the University of Maryland (http://hpcc.umd.edu).  We acknowledge extensive use of NASA's Astrophysics Data System (ADS) bibliographic services and arXiv.  We are grateful to Cecilia Chirenti, Alex Dittmann, Isiah Holt, Fred Lamb, D{\'e}bora Mroczek, Jaki Noronha-Hostler, and Nico Yunes for numerous constructive discussions, and to the NICER team for their outstanding work on the construction and calibration of the instrument.

\bibliography{bibfile}

\end{document}